\documentclass[twocolumn,preprintnumbers,amsmath,amssymb,aps,prb]{revtex4}
\usepackage{graphicx}
\begin{document}

\title{Detecting Depinning and Nonequilibrium Transitions with Unsupervised Machine Learning}
\author{
 D. McDermott$^{1,2}$, 
 C. J. O. Reichhardt$^{1}$, and  C. Reichhardt$^{1}$ 
} 
\affiliation{
$^1$Theoretical Division,
Los Alamos National Laboratory, Los Alamos, New Mexico 87545 USA\\ 
$^2$Department of Physics, Pacific University,
Forest Grove, Oregon 97116, USA\\
}


\date{\today}
\begin{abstract}
  Using numerical simulations of a model disk system,
  we demonstrate that a machine learning generated order parameter
  can detect depinning transitions and different dynamic flow phases 
  in systems
  driven far from equilibrium.
  We specifically consider monodisperse passive
  disks with short range
  interactions undergoing 
 a depinning phase transition 
  when driven 
  over quenched disorder.
  The machine learning derived order parameter identifies 
  the depinning transition as well as different dynamical regimes, 
  such as the transition from a flowing liquid
  to a phase separated liquid-solid state
  that is not readily distinguished with traditional
  measures such as velocity-force curves or Voronoi tessellation.
  The order parameter also
  shows markedly distinct behavior 
  in the limit of high density
  where jamming effects occur.
  Our results should be general to the broad class of particle-based systems
that exhibit depinning transitions and nonequilibrium phase transitions. 
\end{abstract}
\maketitle
    
\section{Introduction}
Principal component analysis (PCA)
is a linear algebra algorithm that 
is widely used
for identifying patterns in data sets \cite{Abdi10}.
PCA 
determines the axes along which a data set
has the largest variance
by expressing the set 
as a linear sum of basis vectors.
The principal components (PC)
are the eigenvectors of the data matrix 
eigenvalue, 
and the first principal eigenvector 
has the largest possible eigenvalue.
PCA is closely related to single vector decomposition,
as detailed in Ref.~\cite{Shlens14},
a primer on the mathematics of PCA. 
Applications of PCA include 
whitening data 
and reducing the dimensionality of the system.
Often researchers
with large data sets
use PCA as a tool 
to reveal a hidden underlying relationship
among variables
by changing the basis in which the data is expressed
and computing
the principal components. 
In reducible data sets,
the sum of basis vectors can be truncated while maintaining
a reasonable approximation of the original data.

 PCA in conjunction with machine learning is 
 used in a broad range of fields 
 where large data sets are common
 and the underlying relationship between
 the variables may not be apparent, such as
 in biology \cite{McKinney06,Pang16} 
 and pattern recognition \cite{Bishop06}. 
 Use of PCA
 requires the construction of a feature matrix,
 where a feature is a general name for a measurement of a system.
 Typically,
 the feature matrix contains
 data trials along its rows, 
 which are often called samples in machine learning, and 
 different features 
 along its columns.
 For example, in facial recognition applications,
 the pixel values of a digitized photo are used as features,
 and each photo is considered a sample.  
 The PCA algorithm fits the data and
 outputs 
 the principal components
 written as linear combinations of the original features.
 The algorithm
 synthesizes information
 using minimization techniques that maximize
 the variance along the principal components,
 which may result from an underlying fundamental
 physical model.

 In condensed matter physics,
 PCA has successfully been applied
 to detect phase transitions
 in
 the Ising and XY models 
 based on
 matrices of raw spin configurations \cite{Carrasquilla17,Wetzel17,Hu17,Wang16}.
 In a detailed study of various spin models,
 Hu {\it et al.} \cite{Hu17} 
 confirmed that
 PCA is well suited for recognizing
 order and symmetry breaking,
 and showed that
 the distribution of principal components
 can be used to
 separate
 strong first order transitions from second order transitions
 as well as to
 distinguish
 phase transitions from pattern changes.
 Equilibrium phase transitions in a variety of
 soft matter systems not confined to a lattice
 can be detected with
 PCA,
 such as a density-driven liquid to hexatic
 phase transition in passive disks \cite{Jadrich18,Jadrich18a},
 where PCA was able to reproduce the
 qualitative shape of the traditional order parameters.
 In Ref.~\cite{Jadrich18a}
 the method was extended
 to non-equilibrium
 phase transitions
 such as random organization \cite{Corte08}. 
 In these studies, the features
 are
 constructed intuitively
 using measures similar to the
 pair correlation function $g(r)$,
 which is known to be an excellent indicator of both
 short and long range order
 in tightly packed particle systems.
 Intriguingly,
 the transformed principal components found by PCA can be related 
 to the packing structures in the disk systems,
 and by modifying the sampling of
 the number
 of probe particles and neighbor particles,
 it
 is possible to develop 
 physical insights regarding the
 shape and magnitude of the order parameter \cite{Jadrich18a}.

 Since PCA methods have proven
 successful at characterizing certain 
 nonequilibrium systems,
 it is natural
 to apply these methods
 in systems that 
 exhibit depinning transitions when
 individual particles, groups of particles, or elastically coupled 
 elements
 are driven over quenched disorder \cite{Corte08,Reichhardt17}.
 Such behavior arises for the depinning of
 magnetic vortex lines in type-II superconductors
 \cite{Bhattacharya93,Koshelev94,Reichhardt97,Olson98a,Kolton99},
 magnetic domain walls \cite{Atkinson03}, contact lines \cite{Paxson13}, 
 electron crystals \cite{Williams91,Cha94a,Reichhardt01},
 stripe and bubble phases, \cite{Cooper03,Reichhardt03a,Zhao13,Wang15}, sliding
 quantum crystals \cite{Brussarski18},
 skyrmions \cite{Schulz12,Nagaosa13,Reichhardt15a,Jiang17,Legrand17,Diaz17},
 sliding charge 
 density waves \cite{Gruner88,Myers93,Li99},
 colloids \cite{Reichhardt02,Pertsinidis08,Tierno12,Bohlein12,Vanossi12,McDermott13},
 jammed systems with
 quenched disorder \cite{Reichhardt12,Graves16},
 sliding friction \cite{Cule96,Tekic05,Vanossi13}, geological systems \cite{Carlson89},
 dislocation dynamics \cite{Miguel02,Zhou15},
 pattern forming systems \cite{Sengupta10},
 and  active matter \cite{Morin17,Sandor17}.
 In addition to the
 depinning transition,
 these systems can
 exhibit a wealth of distinct dynamical flow phases
 along with
 transitions between these phases,
 such as depinning into a 
 disordered liquid \cite{Koshelev94,Olson98a,Kolton99,Reichhardt01,Reichhardt03a,Legrand17,Reichhardt02,Pertsinidis08,Tierno12}
 followed by a
 transition into a moving crystal
 \cite{Bhattacharya93,Koshelev94,Reichhardt97,Reichhardt15a,Diaz17,Gruner88},
 moving smectic \cite{Kolton99,Atkinson03,Reichhardt03a,Zhao13,Bohlein12,Vanossi12,LeDoussal98,Balents98,Pardo98}, or other  moving
 pattern \cite{Reichhardt03a,Zhao13,Zhou15,Sengupta10,Sandor17} at higher drives.
 Traditional methods to characterize these
 systems include the velocity-force curves,
 differential conductivity, structure factor, and Voronoi tessellations \cite{Reichhardt17};
 however,
 there are many cases in which
 the system exhibits dynamics that appear different to the eye but are not distinct
 according to these standard order parameters.
 Thus, the nature of the appropriate order parameter is often not clear.
 There have been some studies using machine learning algorithms
 to detect depinning transitions
 on ferroelectric relaxors 
 using the k-means algorithm~\cite{Jadrich18a}.
 It is, however,
 an open question whether a PCA
 generated order parameter 
 can characterize
 transient and steady state
 nonequilibrium
 flow phases,
 as well as non-equilibrium
 phase transitions, 
 such as those
 observed in particle based systems. 

 In this paper,
 we apply PCA to driven monodisperse disk systems with quenched disorder.
 Despite the apparent simplicity of this system, it
 exhibits not only
 depinning transitions but
 also a variety of distinct dynamical phases,
 including laning, clustering, crystalline, and jammed phases \cite{Yang17}.
 Often the transitions between these phases
 only produce weak signatures in the
 standard order parameters.
 We demonstrate that
 PCA can
 automatically detect the different dynamic behaviors as a function of drive.  
 The features we employ
 are constructed from intuitive measures similar to
 the pair correlation function $g(r)$ used in
 Jadrich {\it et al.} \cite{Jadrich18,Jadrich18a} 
 We show that
 the machine learning derived order parameter 
 is superior
 to the standard order parameters,
 indicating that combining the pairwise distance information
 into principal components 
 using PCA
 can successfully synthesize the fundamental information
 of the emergent behavior.
 This method
 could be applied to a wide variety of other particle-based systems
 that exhibit depinning.

 The paper is organized as follows.
 In Section II we outline the
 principal component analysis
 technique for the depinning system.
 The simulation details of the disk system, along with the standard
 measures such as the velocity-force curve and Voronoi tessellation,
 are described
 in Section III.
 We show in Section IV that the
 principal component analysis of the disk system at
 different densities
 can identify distinct changes which correlate with
 changes in the dynamics and structure of the system,
 and in Section V we summarize our results.   

 \section{Principal Component Analysis of Disk Systems}
 \label{sec:PCA}  
 PCA is designed to discover and maximize correlations in data sets
 contained in matrices ~\cite{Hu17}.
 The features range from
 pixel values of the digitized photo
 in facial recognition applications
 to the matrix of spin values
 in spin-based systems.
 In off-lattice systems,
 the raw position data does not naturally lend itself
 to description by an $m$ by $n$ matrix.
 Thus to apply PCA to disks, it is useful to apply
 traditional measures of
 structural information.
 Here we use a ``particle centered'' measure
 in PCA to perform dimensionality reduction
 on the geometrical environment of the particles themselves, 
 rather than
 attempting to classify the manner in which particles fit into
 the container. 
 We consider a 2D system of disks of radius $r$ interacting
 with a random  array of pinning sites, as described in \cite{Yang17}; additional
 simulation details appear in Section III.

 We characterize the structural information
 of the disks using 
 the relative positional data, $r_{ij} = |{\bf r}_{ij}|$,
 where ${\bf r}_{ij}={\bf r}_i-{\bf r}_j$ is the center-to-center distance
 between disks $i$ and $j$.
 For a certain subset of
 probe particles $m=N_{\rm probe}=1000$,
 we measure the distance from
 the probe particle
 to $n$ of its neighbor particles.
 We sort the resulting distances and place the values into a feature vector
 for each probe particle $i$,
 \begin{equation}
   \vec{f}_i = [r_{i0},r_{i1},r_{i2}, ..., r_{ij}, ..., r_{in}].
 \end{equation}
 such that $r_{i0} < r_{i1} <  ... < r_{in}$.
 In a procedure typical for PCA, 
 we center
 the feature vectors
 by computing
 the average of each neighbor distance,
 \begin{equation}
   \langle r_j \rangle _D = \frac{1}{N_{\rm probe}}\Sigma_{i=0}^{N_{\rm probe}}{r_{ij}},
 \end{equation}    
 in order to create a vector containing a series of averages:
 \begin{equation}
   \vec{\langle r \rangle}_D = [\langle r_0 \rangle _D, \langle r_1 \rangle _D, ... , \langle r_j \rangle _D, ..., \langle r_n \rangle _D].
 \end{equation}
 We subtract $\vec{\langle r\rangle}_D$ from each feature vector $\vec{f}_i$.

 We assemble the centered feature vectors into a feature matrix:
 \begin{equation}
   \vec{F} = [\vec{f}_{0},\vec{f}_{1},\vec{f}_{2}, ..., \vec{f}_{m}]^T .
 \end{equation}
 To remove the correlations introduced in the sorting process,
 we perform the essential step of
 whitening the data, as in Ref~\cite{Jadrich18,Jadrich18a}. 
 The whitening transformation is performed
 by applying
 the PCA algorithm
 to a feature matrix of an ideal gas composed of 
 non-interacting disks at density $\phi$
 using the same number of probe $N_{\rm probe}$ and neighbor $n$ disks
 as in our actual system.
 The PCA analysis of the ideal gas system generates
 a transformation matrix, $\vec{W}_0(\phi)$,
 that transforms the ideal gas system with density $\phi$
 into a Gaussian distribution with mean zero and variance one.
 We can then use
 $\vec{W}_0$
 to transform
 feature matrices from 
 non-ideal gas systems with density $\phi$
 into a coordinate space in which 
 naive sorting correlations have been removed,
 preserving only the features that contain
 correlations
 due to the particle interactions and external forces.
 It is necessary to compute a separate $\vec{W}_0$ for each density $\phi$.

 To implement PCA, we use the incremental PCA
 library available through Scikit-Learn \cite{Pedregosa11}
 in order to process many feature vectors
 without holding the entire feature matrix in memory.
 After analyzing the whole data set, 
 the PCA algorithm returns a transformation matrix, $\vec{W}(\phi)$,
 that can be applied to new data.
 Here, we generate $\vec{W}$ at fixed density $\phi$ values
 for all values of $F_D$.
 Thus the algorithm
 simultaneously ``sees'' systems
 above and below the depinning transition
 in the variety of
 different phases that are described in Ref.~\cite{Yang17}.

 In order to apply the PCA algorithm,
 we run the disk simulation until the system reaches a steady state.
 Then we sample snapshots of the system,
 using ten frames
 spaced by $\Delta t = 1 \times 10^5$ simulation time steps.
 In each frame we randomly
 select $m = 1000$ probe particles,
 and for each particle we calculate the distance to its $n$ nearest neighbors.
 Here we take $n=N_d(\phi)-1$, where
 $N_d(\phi)$ is the total number of disks in the sample
 at density $\phi$, meaning that we calculate the distance from the probe particle
 to all other particles in the system.
 For each simulation frame we generate the
 centered feature vector $\vec{f}_i$.
 This must be prewhitened to obtain $\vec{f}_i^w=\vec{W}_0(\phi)\vec{f}_i$.
 To generate a PCA transformation matrix $\vec{W}(\phi)$ valid for disk density $\phi$, 
 we analyze all feature data from all frames
 for the given disk density $\phi$ by
 feeding ten $m \times n$ matrices of prewhitened feature vectors
 into the incremental PCA algorithm
 in sequence.
 The algorithm returns the transformed data $\vec{f}'$,
 eigenvalues $\lambda_N$,
 and the transformation matrix
 $\vec{W}$.
 This matrix may be used
 to transform new prewhitened feature vectors from subsequent
 snapshots of data obtained at the same value of $\phi$, 
 or it can be applied to the already processed feature vectors
 in order to generate a visualization of the vectors
 in the new basis space.
    
 Jadrich {\it et al.} \cite{Jadrich18,Jadrich18a} 
 showed that
 the principal components
 contain structural information of the system,
 and thus can be used as an order parameter (OP)
 of the system.
 To construct such an order parameter,
 we 
 transform
 a prewhitened feature vector $\vec{f_i}^w$
 obtained at fixed $F_D$ and $\phi$ with the trained PCA model to obtain
 \begin{equation}
   \vec{p}_i = \vec{W}\vec{f}^w_i. 
 \end{equation}
 The order parameter $P_1$ is defined
 to be the extent to which
 the first principal component
 captures the information content in the system,
 \begin{equation}
   P_1 = \langle |p_1| \rangle,
 \end{equation}
 where $p_1$ is the first element in $\vec{p}$.
 We analyze 
 the eigenvalue spectrum using a scree plot
 to
 determine how well the matrix
 can be expressed in the new PC basis.
 We also plot
 the magnitude of the first principal component, 
 the total transformation matrix $\vec{Q}=\vec{W}\vec{W}_0$
 that is remarkably similar to $g(r)$,
 and the ML derived order parameter
 $P_1 = \langle |p_1| \rangle$.
  
\section{Simulation and System}
\label{sec:simulation}
  We analyze the data from our previous publication \cite{Yang17}, 
  in which we performed 2D molecular dynamics simulations of 
  passive disk systems.
  The system contains $N_d$ disks of radius $R_d$
  within a simulation box of $S_x = S_y = 60.0$,
  in dimensionless simulation length units,
  with periodic boundary conditions.
  The area density is given by $\phi = N_d \pi R_d^2/(S_x S_y)$.
  In the absence of quenched disorder,
  the disks form a polycrystalline state near
  $\phi \approx 0.85$ 
  and a triangular solid at $\phi \approx 0.9$.

  The disk dynamics are governed by
  the following overdamped equation of motion: 
  \begin{equation}
    \eta \frac{d {\bf r}_{i}}{dt} = {\bf F}_{dd}  + {\bf F}_{p}  + {\bf F}_{D} .
  \end{equation}
  Here $\eta$ is the damping constant 
  and ${\bf r}_{i}$ is the location of disk $i$. 
  The disk-disk interaction force is 
  ${\bf F}_{dd} = \sum_{i\neq j}k(2R_{d} - |{\bf r}_{ij}|)\Theta(2R_{d} - |{\bf r}_{ij}|) {\hat {\bf r}_{ij}}$,
  where ${\bf r}_{ij} = {\bf r}_{i} - {\bf r}_{j}$,  $\hat {\bf r}_{ij}  = {\bf r}_{ij}/|{\bf r}_{ij}|$,
  the disk radius $R_{d} = 0.5$, and
  the spring constant $k = 50$.
  Distances are measured in simulation units
  $l_0$ and forces are measured in simulation units $f_0$ so
  that $k$ is in units of $f_0/l_0$ and the unit of simulation time is $\tau=\eta l_0/f_0$.

  We introduce quenched disorder by
  placing pinning sites
  throughout the sample.
  The pinning force ${\bf F}_p$
  is modeled as arising from $N_p$ randomly placed parabolic attractive 
  wells with a pinning radius of $r_{p} = 0.5$,
  such that only a single disk
  can be trapped in a given pinning site at any given time.
  We fix the pinning density to
  $\phi_p = N_p/(S_x S_y) = 0.314$.
  The driving force ${\bf F}_D=F_D{\bf \hat{x}}$ is applied uniformly to
  all particles and is
  incremented in steps of $\Delta F_D = 0.05$
  after every $\Delta t = 1 \times 10^6$ simulation time steps.
  At each drive increment, we measure 
  the average disk velocity
  $\langle V_{x}\rangle = N_d^{-1}\sum^{N_d}_{i=1}{\bf v}_{i}\cdot {\hat{\bf x} }$,
  where ${\bf v}_{i}$ is the
  instantaneous velocity of disk $i$.

  A useful measure 
  for characterizing
  interacting particles driven over disorder is the fraction
  $P_6$ of sixfold-coordinated particles.
  Here $P_6=N_d^{-1}\sum_{i}^{N_d}\delta(z_i-6)$,
  where $z_i$ is the coordination
  number of disk $i$ obtained from a Voronoi tessellation.
  Previously \cite{Yang17},
  we correlated local maxima in $P_6$
  with changes in the dynamic phases,
  and demonstrated that $P_6$
  did not have a feature at all of the dynamical phase transitions.
  This behavior for the disks with short-range interactions
  differs from what is observed for particles
  that have longer range interactions,
  where the dynamic phase changes are more readily detected using
  information from the Voronoi tessellation.

  \begin{figure}
    \includegraphics[width=3.5in]{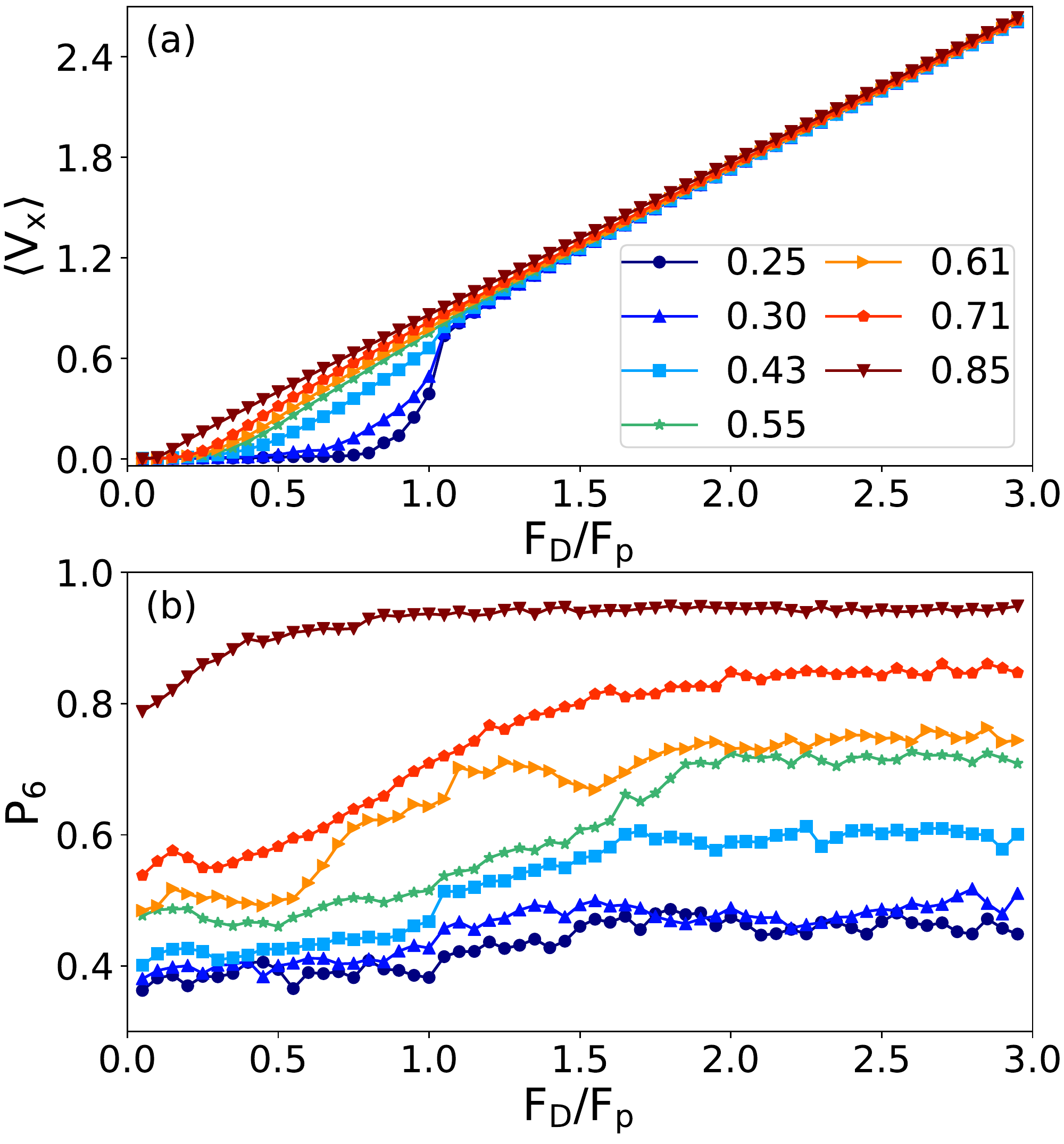}
    \caption{
      (a) The 
      average disk velocity
      $\langle V_{x} \rangle$ 
      vs driving force
      $F_{D}/F_{p}$ in samples with 
      total 
      disk density $\phi$ of
      $\phi=0.85$ (down triangles), 
      $0.71$ (pentagons), 
      $0.61$ (right triangles), 
      $0.55$ (stars),   
      $0.43$ (squares), 
      $0.30$ (up triangles), 
      and 
      $0.25$ (circles). 
      (b) The corresponding $P_6$ vs $F_D/F_p$.
    }
    \label{fig:1}
  \end{figure}
  
  \section{Results}
  \label{sec:1}
  In Fig.~\ref{fig:1}
  we plot the traditional
  dynamical measurements as a function of $F_D/F_p$ 
  for a sample with fixed pinning
  density at different disk densities $\phi$ ranging from $\phi=0.25$ to $\phi=0.85.$
  The velocity-force curves $\langle V_x\rangle$ versus $F_D/F_p$ in 
  Fig.~\ref{fig:1}(a) have the same features that are generically found in
  systems that undergo depinning.
  At low drive,
  there is a pinned regime with $\langle V_{x}\rangle = 0.$
  This is followed at higher drive by
  a nonlinear regime above depinning,
  and at the highest drives, there is a regime
  in which the velocity increases linearly 
  with increasing $F_{D}$.
  As the disk density increases,
  the depinning transition shifts to lower $F_D$ and the region of
  nonlinear velocity response becomes narrower.
  In Fig.~\ref{fig:1}(b),
  the fraction $P_6$ of six-fold coordinated disks versus $F_D/F_p$
  is nearly flat
  for $\phi =  0.25$ and $0.3$, while for higher $\phi$ there is some tendency
  for $P_{6}$ to increase with increasing $F_{D}$. 
  Overall, the results in Fig.~\ref{fig:1} indicate
  that it is
  difficult to
  identify distinct dynamic phases using these measures,
  and that it is even difficult
  to precisely pinpoint the depinning transition.

  \begin{figure}
    \includegraphics[width=3.5in]{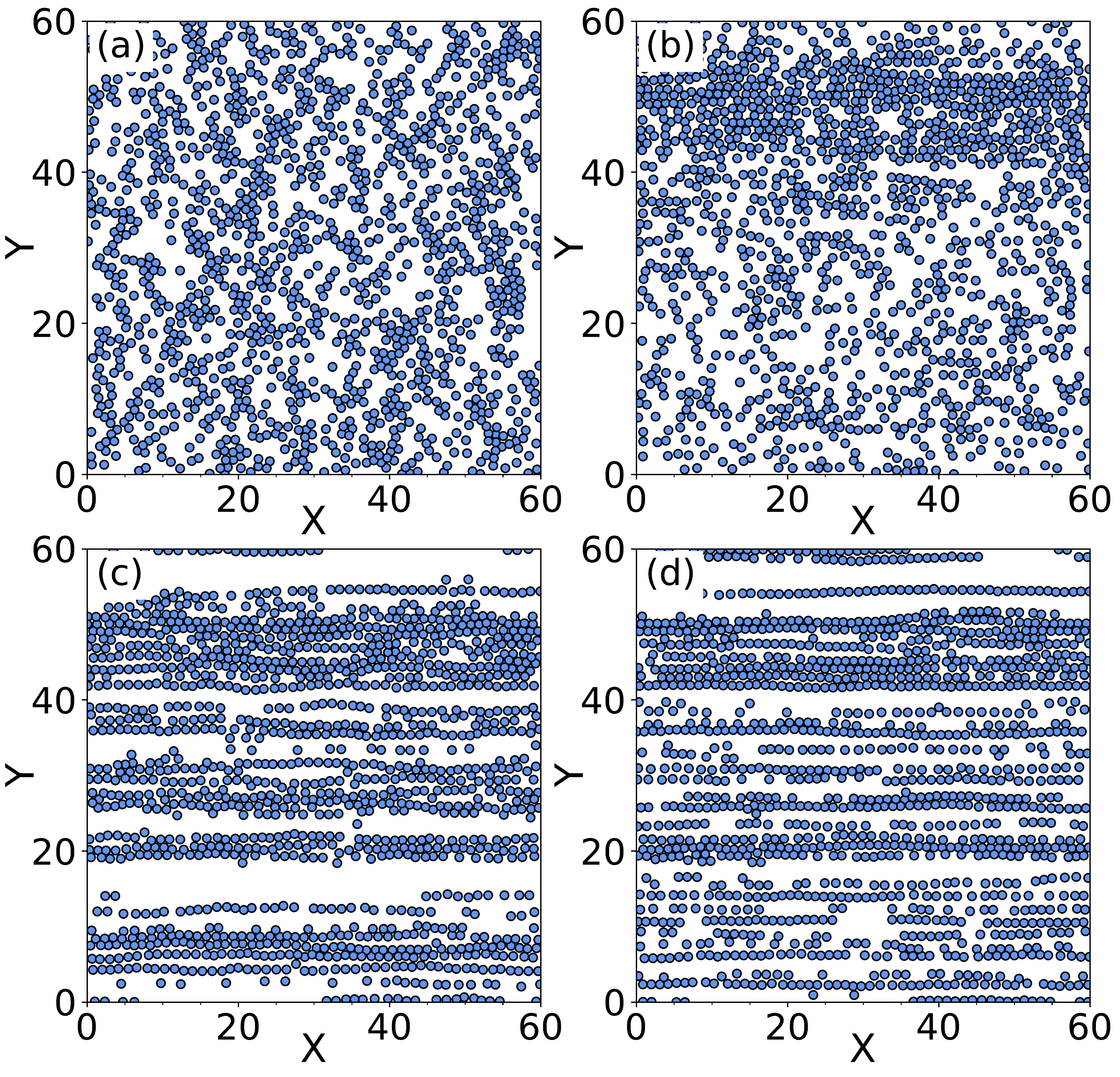}
    \caption{
      Images obtained at a disk density of $\phi=0.30$ for the system in Fig.~\ref{fig:1}
      at $F_{D} = 0.25$ (pinned),
      (b) $F_{D} = 0.95$ (phase separated),
      (c) $F_{D} = 1.5$ (smectic flow), and
      (d) $F_{D} = 2.5$ (smectic flow).   
    }
    \label{fig:2}
  \end{figure}

\subsection{Low Disk Density}
  We first focus on the low density limit with $\phi=0.25$ and $\phi=0.3$,
  where $P_{6}$ is almost featureless.
  In Fig.~\ref{fig:2} we illustrate the disk configurations
  in a sample with 
  $\phi = 0.3$ at different values of $F_D/F_p$.
  In the pinned phase at $F_D/F_p=0.25$, Fig.~\ref{fig:2}(a) 
  shows that 
  small disordered clusters appear since
  some of the particles have formed clogged clusters instead of being
  directly trapped by the pinning sites.
  Above depinning at $F_D/F_p=0.95$ in Fig.~\ref{fig:2}(b),
  there is a combination of smaller pinned clusters
  with a phase separated
  region of higher density in which the disks move in a band.
  In Fig.~\ref{fig:2}(c)
  at $F_{D}/F_p = 1.5$,
  all the disks are moving
  in one-dimensional (1D) chains,
  while in Fig.~\ref{fig:2}(d) at $F_{D} = 2.5$,
  the moving chains have become somewhat more rarefied.
  These results indicate that
  different dynamical regimes are present which
  are generally not detectable with the standard measures.
  We note that other measures such as
  the structure factor $S(k)$ and diffusion
  similarly show only weak or no changes at the transitions among these
  dynamical regimes.

 \begin{figure}
  \includegraphics[width=0.45\textwidth]{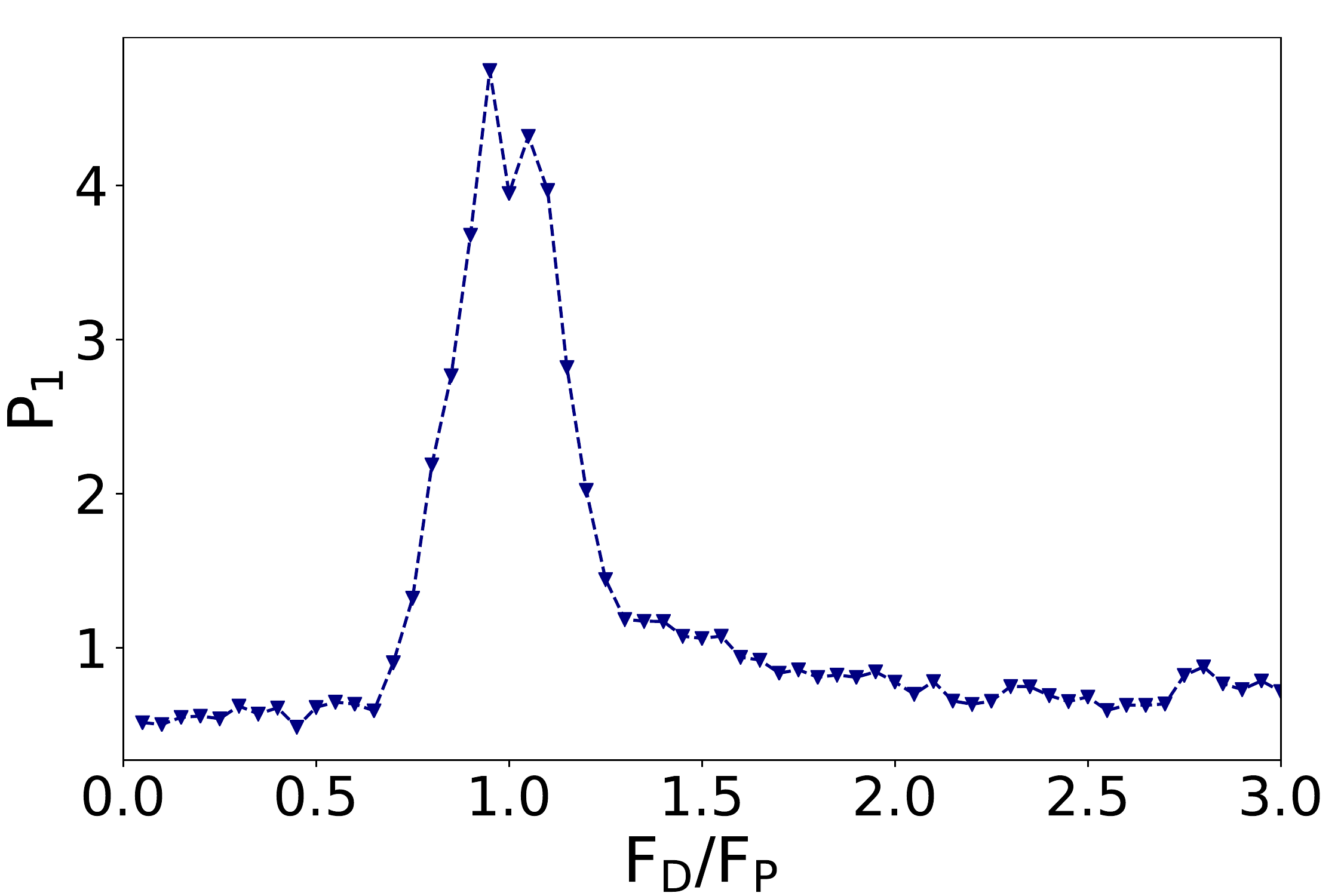} 
  \caption{
    The machine learning derived order parameter $P_1$
    vs $F_D/F_p$
    for the system in Fig.~\ref{fig:1}
    at disk density $\phi=0.30$.
    Three phases are clearly apparent:
    the pinned state at $0 < F_{D}/F_p < 0.65$, 
    the phase separated
    state at $0.65 \leq F_{D}/F_p < 1.3$,
    and the smectic or laned state at $F_{D} \geq 1.3$. 
  }
  \label{fig:3}
 \end{figure}
  
 In Fig.~\ref{fig:3}
 we plot the machine learning
 derived order parameter $P_1$ versus $F_{D}/F_p$
 for the system in Fig.~\ref{fig:1}.
 We find 
 $P_1 \approx 0.5$
 for $0 < F_D/F_p < 0.65$,
 which is the pinned state illustrated in Fig.~\ref{fig:2}(a).
 This is followed by an increase in $P_1$ 
 at $F_{D}/F_{p} = 0.65$, corresponding to the depinning transition.
 $P_1$ remains elevated over the range
 $0.65 \leq F_D/F_p < 1.05$
 in the phase separated state shown in Fig.~\ref{fig:2}(b).
 For $1.05 \leq F_D/F_p < 1.25$,
 $P_1$ decreases
 when the system crosses over
 into the smectic or laned state.
 There is a gradual decrease in $P_1$ from
 $P_1\approx 1.2$ to $P_1 \approx 1.0$
 over the range $1.25 \leq F_D/F_p \leq 3.0$
as the smectic lanes become increasingly well defined,
 as shown
 in Fig.~\ref{fig:2}(c,d)
 at  $F_D/F_p = 1.5$ and $F_{D}/F_{D} = 2.5$. 
 The results in Fig.~\ref{fig:3} indicate that $P_{1}$ clearly detects and
 distinguishes the three phases,
 pinned, phase separated, and smectic,
 along with the transitions between these states.
 For $\phi=0.25$ (not shown), we find similar phases and a similar response of $P_1$.

 \begin{figure}
   \begin{center}
     \includegraphics[width=3.5in]{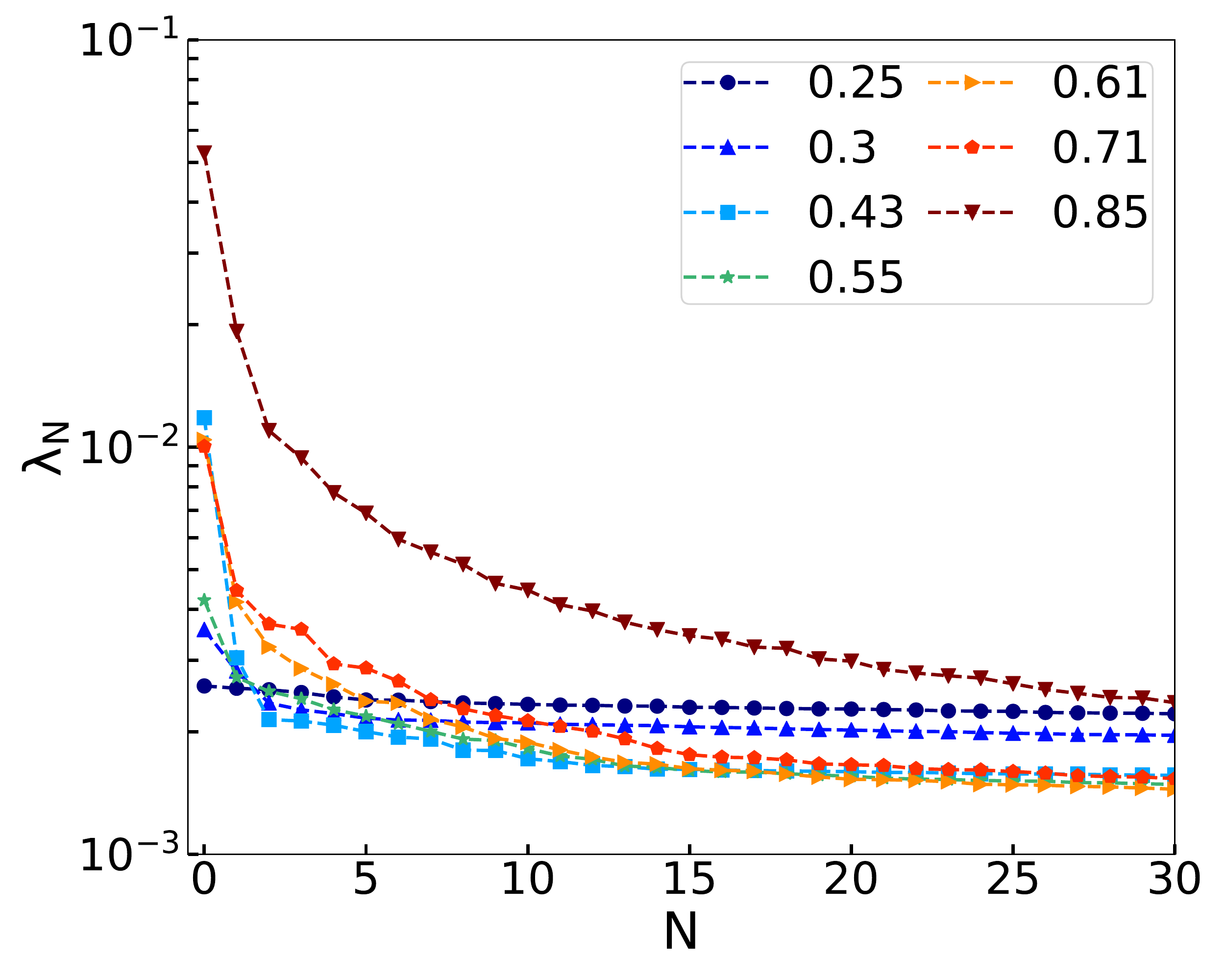}
   \end{center}
   \caption{
     The scaled eigenvalues (relative scores) $\lambda_N$ of the PCA algorithm
     vs the relative ranking $N$ for disk densities of
     $\phi=0.85$ (down triangles), 
     $0.71$ (pentagons), 
     $0.61$ (right triangles), 
     $0.55$ (stars),   
     $0.43$ (squares), 
     $0.30$ (up triangles), 
     and 
     $0.25$ (circles). 
     $N=1$ corresponds to the first principal component.
    }
    \label{fig:4}
  \end{figure}

\subsection{Eigenvalue Distribution} 
  In Fig.~\ref{fig:4}
  we show a scree plot of the eigenvalues $\lambda_N$
  for the samples in
  Fig.~\ref{fig:1} with disk densities
  of $\phi=0.25$ to $\phi=0.85$.
  Here the eigenvalues are sorted from largest ($N=1$) to smallest
  and plotted versus eigenvalue ranking $N$.
  The scree plot gives an indication of how successfully the PCA
  has reduced the dimensionality of the information present in
  the system.
  When the
  eigenvalue spectrum is dominated by 
  one or a few large values of low rank,
  followed by many small values,
  it indicates that the first few eigenvectors
  can be used to describe the primary characteristics
  of the system, since 
  a linear combination
  of the first few principal components
  captures most of the information.
  At $\phi = 0.25$ and $\phi=0.3$,
  the first eigenvalue $\lambda_1$ is 
  somewhat larger in size and the remainder of the eigenvalues are nearly flat.
  At intermediate disk densities of
  $\phi=0.43$ to $\phi=0.73$,
  $\lambda_1$ is substantially larger than
  the remaining eigenvalues,
  indicating that the PCA analysis has captured the features of
  the system well.
  We find
  a significant jump up in all the eigenvalues at the high density of
  $\phi = 0.85$,
  which corresponds to the onset of jamming behavior.

\subsection{Intermediate Disk Densities}
     
 \begin{figure}[h!]
  \includegraphics[width=0.45\textwidth]{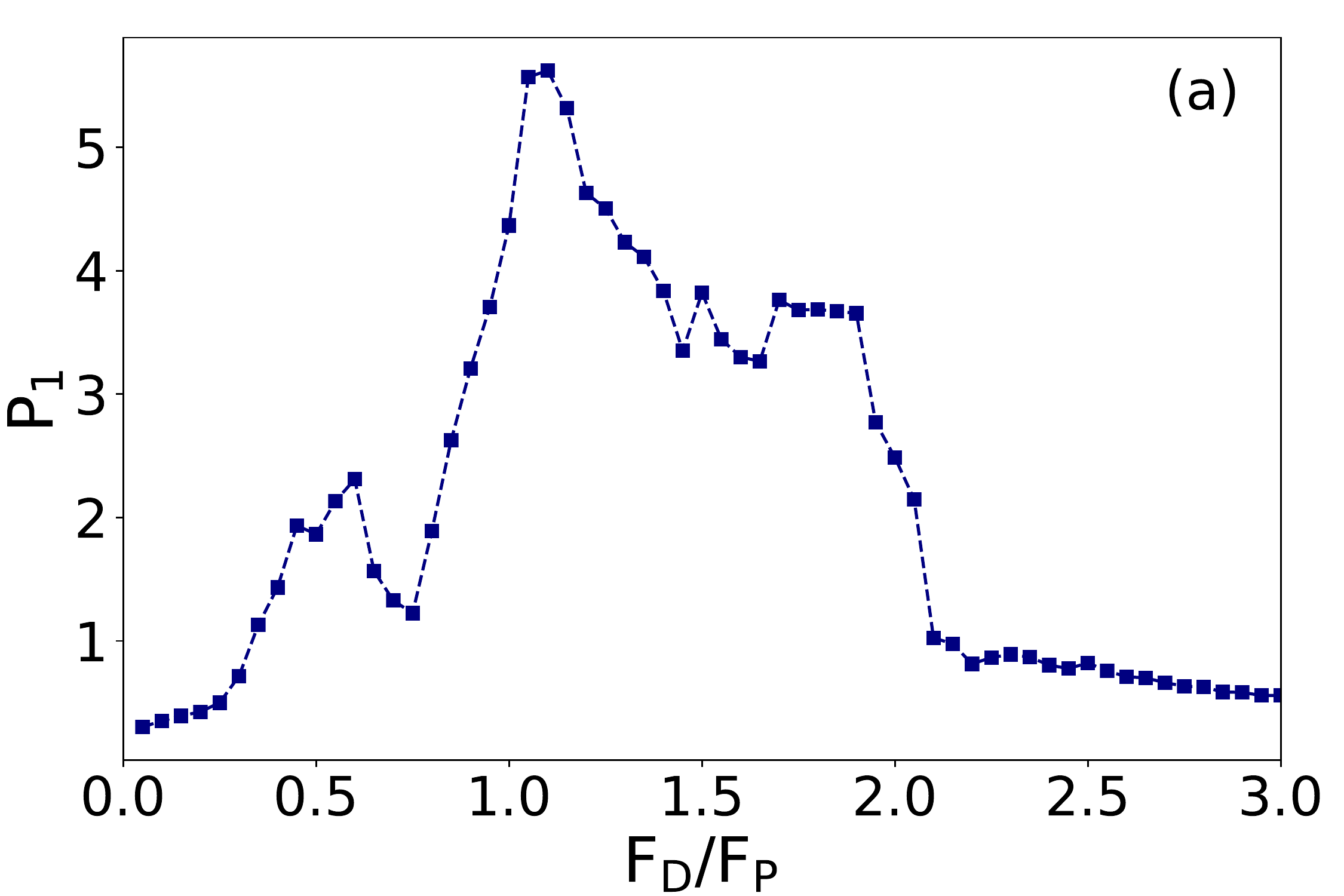}
  \includegraphics[width=0.45\textwidth]{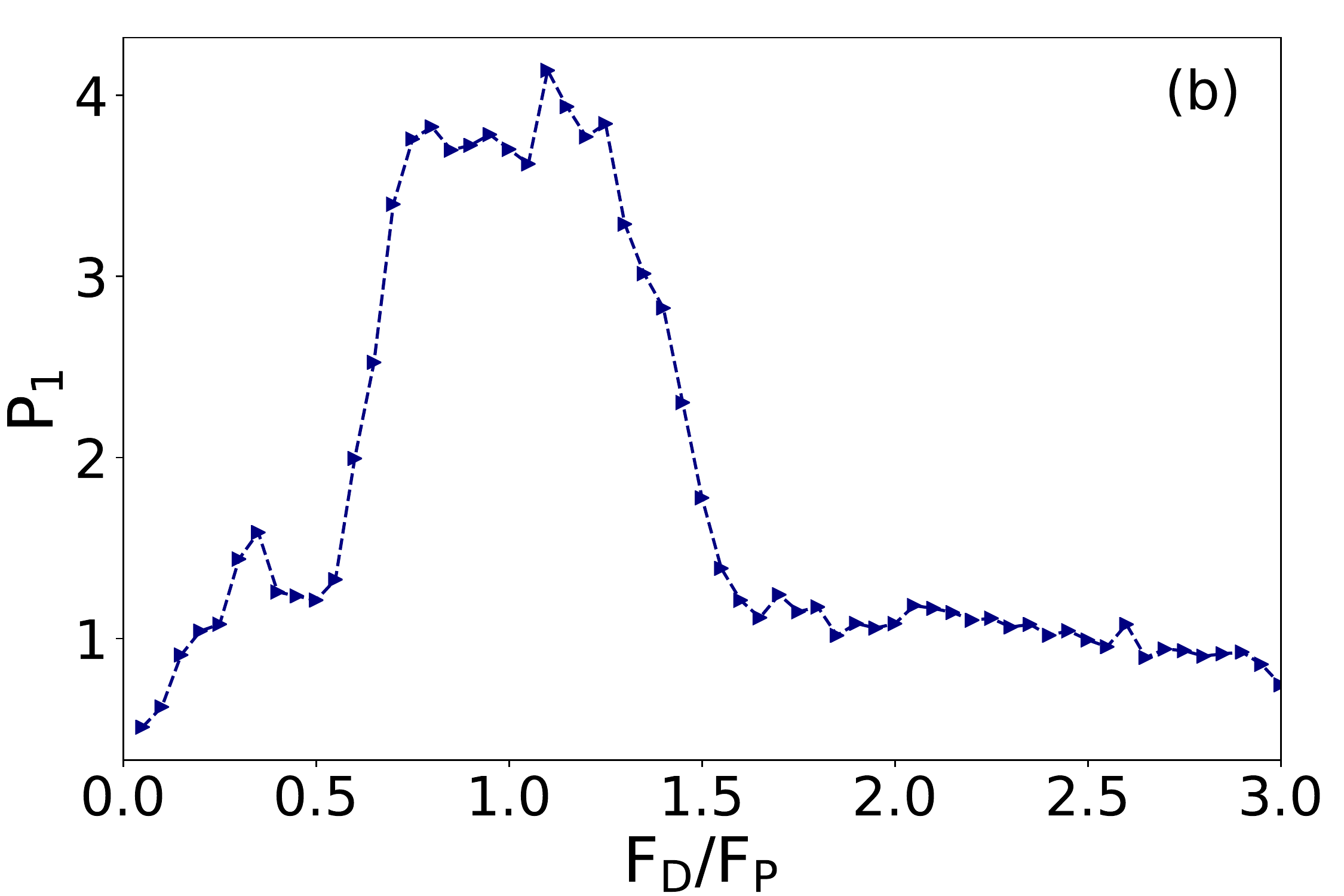}
  \caption{
    $P_1$ vs $F_D/F_p$
    for the system in Fig.~\ref{fig:1} at intermediate disk densities.
    (a) At $\phi = 0.43$, there are multiple peaks.
    (b) At $\phi= 0.61$, the peak structure is more compressed.   
  }
  \label{fig:5}
 \end{figure}

 \begin{figure}
   \includegraphics[width=3.5in]{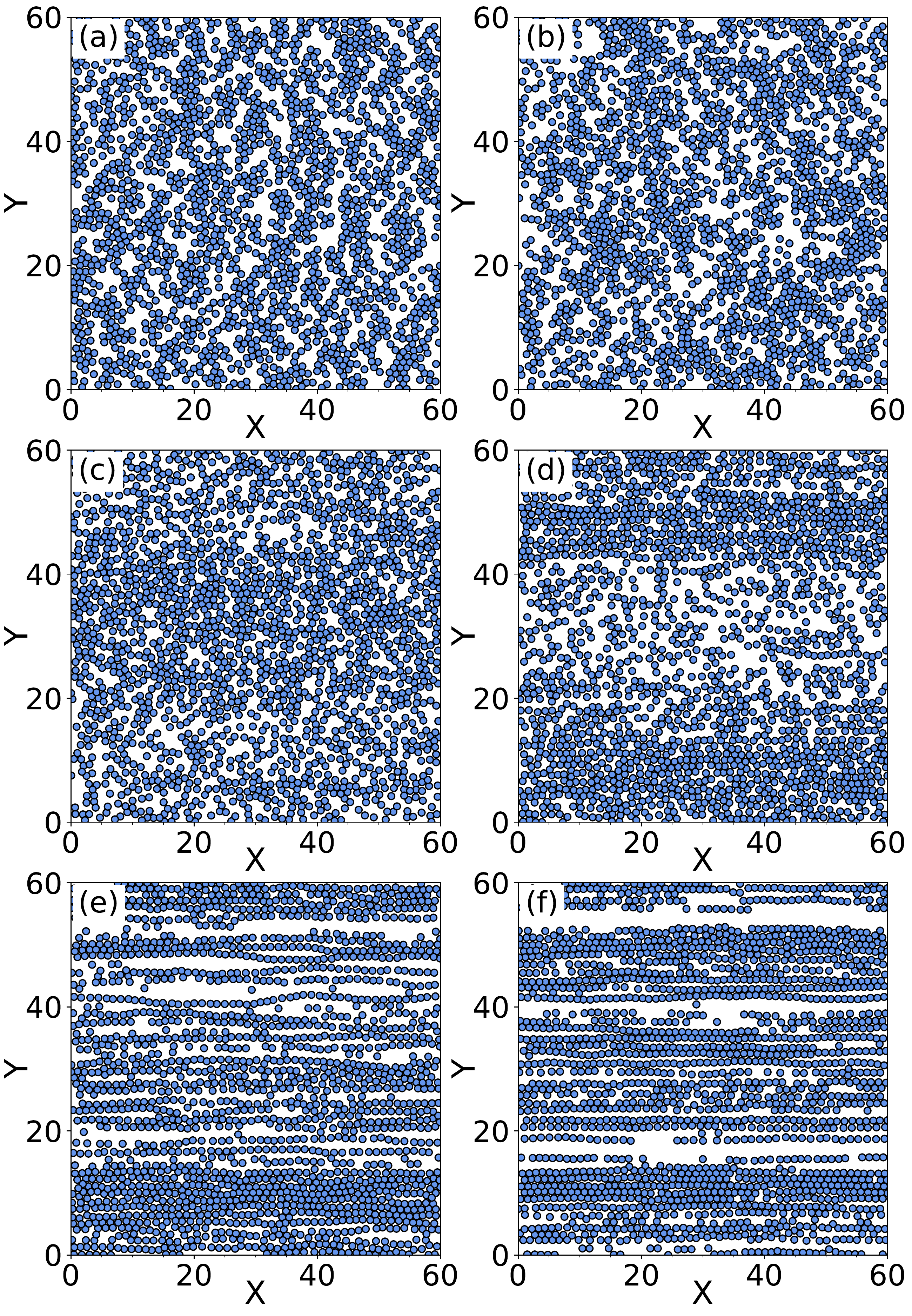}
   \caption{
     Images  obtained at a disk density of $\phi=0.43$ for the system in Fig.~\ref{fig:5}(a) at
     (a) $F_{D}/F_{p} = 0.05$ (disordered pinned state), (b)
     $F_{D}/F_{p} = 0.55$ (clustering clogged state),
     (c) $F_{D}/F_{p}= 0.75$ (moving liquid),
     (d) $F_{D}/F_{p} = 0.95$ (phase separated state with amorphous order), 
     (e) $F_{D}/F_{p} = 1.5$ (phase separated state with crystalline order), and 
     (f) $F_{D}/F_{p} = 2.5$ (moving smectic).    
   }
   \label{fig:6}
 \end{figure}

 We next consider the intermediate disk density regime.
 In Fig.~\ref{fig:5}(a) and (b) we plot $P_{1}$ versus $F_{D}/F_p$ for 
 the samples from Fig.~\ref{fig:1} with $\phi = 0.43$ and $\phi=0.61$, respectively.
 At both densities, in Fig.~\ref{fig:1} $\langle V_{x}\rangle$ versus $F_D/F_p$
 is fairly smooth
 and $P_{6}$ has a gradual increase, 
 but
 it is difficult to 
 distinguish different phases from these measures. 
 In contrast, $P_1$
 in Fig.~\ref{fig:5}(a) $P_{1}$
 has two clear peaks
 at $F_{D}/F_{p} = 0.55$ and
 $F_D/F_p=1.0$, a plateau region over the range $1.5 < F_{D} < 2.0$, 
 and drops to a low value for $F_{D} > 2.0$.
 In Fig.~\ref{fig:6}(a)  we illustrate
 the disk configuration at $F_{D}/F_{p} = 0.05$ within the pinned phase,
 where $P_{1}$ in Fig.~\ref{fig:5}(a) is small.
 Here the disks form small clusters in the pinned state.
 At $F_{D}/F_{p} = 0.55$ in Fig.~\ref{fig:6}(b), just below the
 depinning transition,
 the disks form a locally clustered or clogged state,
 and at depinning these clusters partially break apart,
 producing
 the dip in $P_{1}$ found in Fig.~\ref{fig:5}(a).
 A local minimum in $P_1$ appears near $F_{D}/F_{p} = 0.75$, where the
 structure is a moving liquid
 as shown in Fig.~\ref{fig:6}(c).
 The amorphous phase separated state at
 $F_{D}/F_{p} = 0.95$
 is illustrated in Fig.~\ref{fig:6}(d).
 At $F_{D}/F_{p} = 1.5$, the system is still phase separated
 but the amount of crystalline ordering has increased.
 In Fig.~\ref{fig:6}(f) the configuration at
 $F_{D}/F_{p} = 2.5$ indicates that the disks
 have formed a
 moving smectic state.
 In general, $P_{1}$ shows a pronounced drop at the transition
 into the moving smectic states,
 while the corresponding $P_6$ curve for
 $\phi=0.43$ in Fig.~\ref{fig:1}
 exhibits no feature
 near $F_{D}/F_{p} = 2.5$. 
 This indicates that $P_{1}$ is much more sensitive to the 
 changes in the disk configurations than
 $P_{6}$ or $\langle V_{x}\rangle$.

 \begin{figure}
   \includegraphics[width=3.5in]{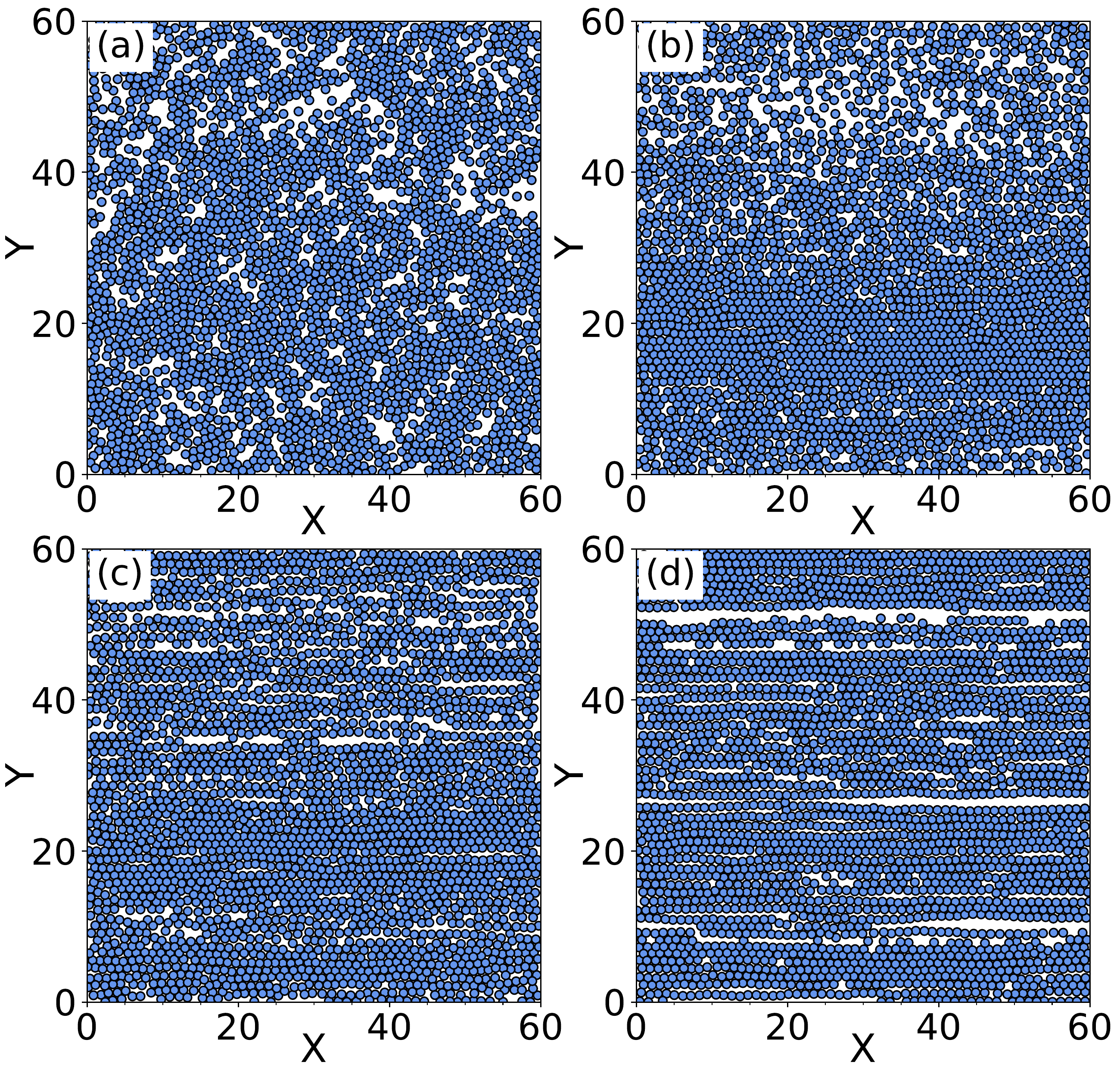}
   \caption{
     Images obtained at a disk density of
     $\phi=0.61$ for the system in Fig.~\ref{fig:5}(b) at
     (a) $F_{D}/F_{p} = 0.25$ (pinned clogged state), (b)
     $F_{D}/F_{p} = 0.95$, (phase separated with crystalline order),
     (c) $F_{D}/F_{p}= 1.5$ (moving state),
     and (d) $F_{D}/F_{p} = 2.5$ (moving state).
   }
   \label{fig:7}
 \end{figure}
 
 For $\phi = 0.61$, Fig.~\ref{fig:5}(b) shows that
 $P_{1}$ versus $F_{D}/F_{p}$ has a similar trend
 as that found for $\phi = 0.43$. There are some differences, however;
 the plateau region in $P_{1}$ is smaller for $\phi=0.61$
 and the drop in $P_{1}$ has shifted to
 a lower value of $F_{D}/F_{p} = 1.5$.
 In Fig.~\ref{fig:7}(a) we illustrate the disk configuration for the
 $\phi=0.61$ system at
 $F_{D}/F_{p} = 0.25$, where
 a pinned clogged state appears.
 This is the same value of $F_D/F_p$ at which
 there is a local peak 
 in $P_{1}$.
 Figure~\ref{fig:7}(b) shows the
 disk configuration at $F_{D}/F_p = 0.95$, where the system
 forms a phase separated state with local crystalline ordering.
 Here $P_{1} = 3.85$,
 which is close to the same value
 found for $P_{1}$ in the $\phi = 0.43$ sample in the
 phase separated moving crystal phase illustrated in Fig.~\ref{fig:6}(e) at $F_D/F_p=1.5$.
 Thus, at $\phi=0.61$,
 the moving phase separated amorphous state found at lower $\phi$
 is missing.
 In Fig.~\ref{fig:7}(c) 
 we show the disk configuration at $F_{D}= 1.5$,
 where the local phase separation is reduced and the system
 begins to form a moving state.
 This moving state becomes more pronounced in Fig.~\ref{fig:7}(d) at $F_{D} = 2.5$.
 
\subsection{High Disk Densities}  

  \begin{figure}[h!]
    \includegraphics[width=0.45\textwidth]{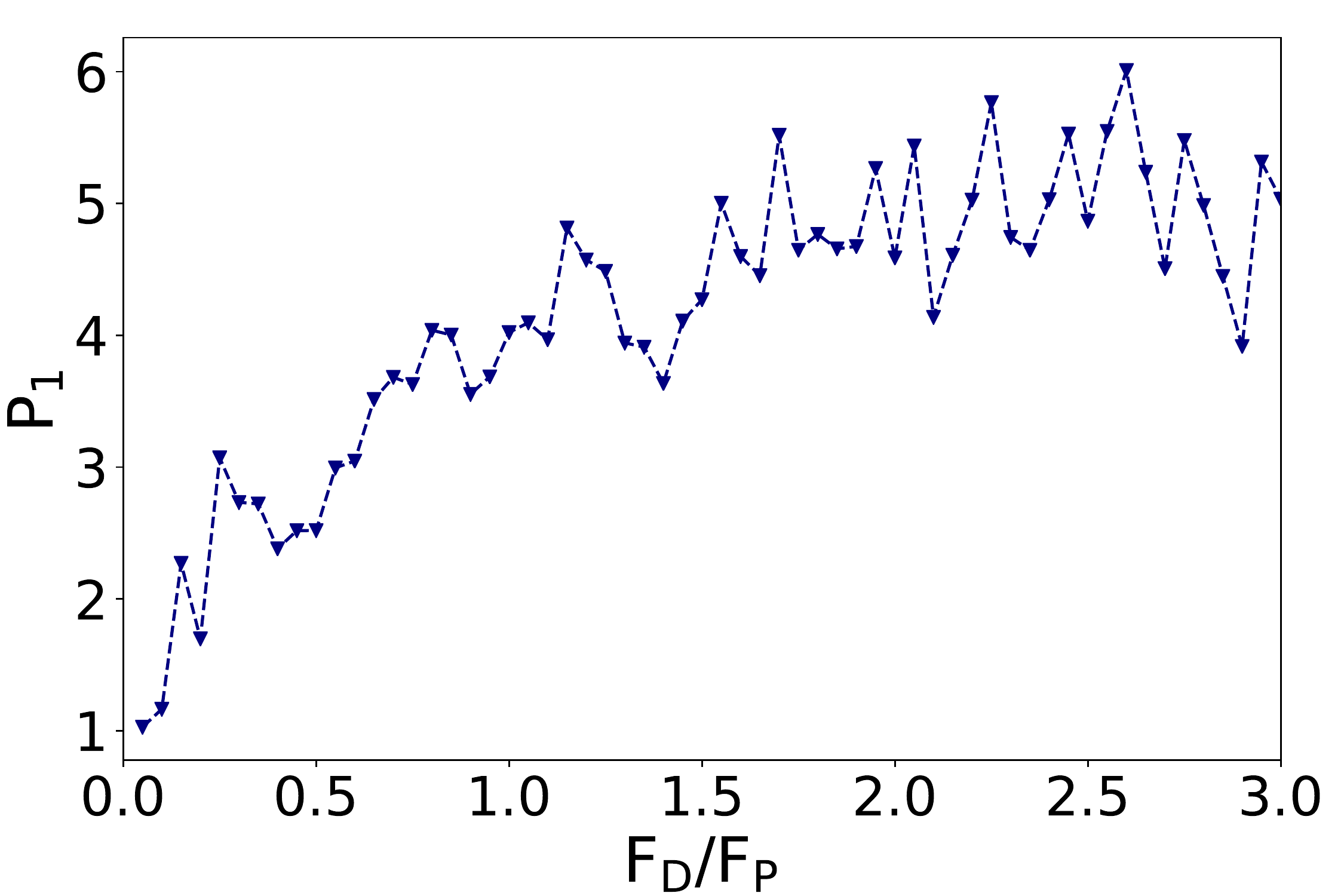}
    \caption{
      $P_1$ vs $F_{D}/F_{p}$ for the system in Fig.~\ref{fig:1} at
      $\phi = 0.85$, where the disks exhibit jamming behavior
      and elastic depinning.  
    }
    \label{fig:8}
  \end{figure}

  \begin{figure}
    \includegraphics[width=3.5in]{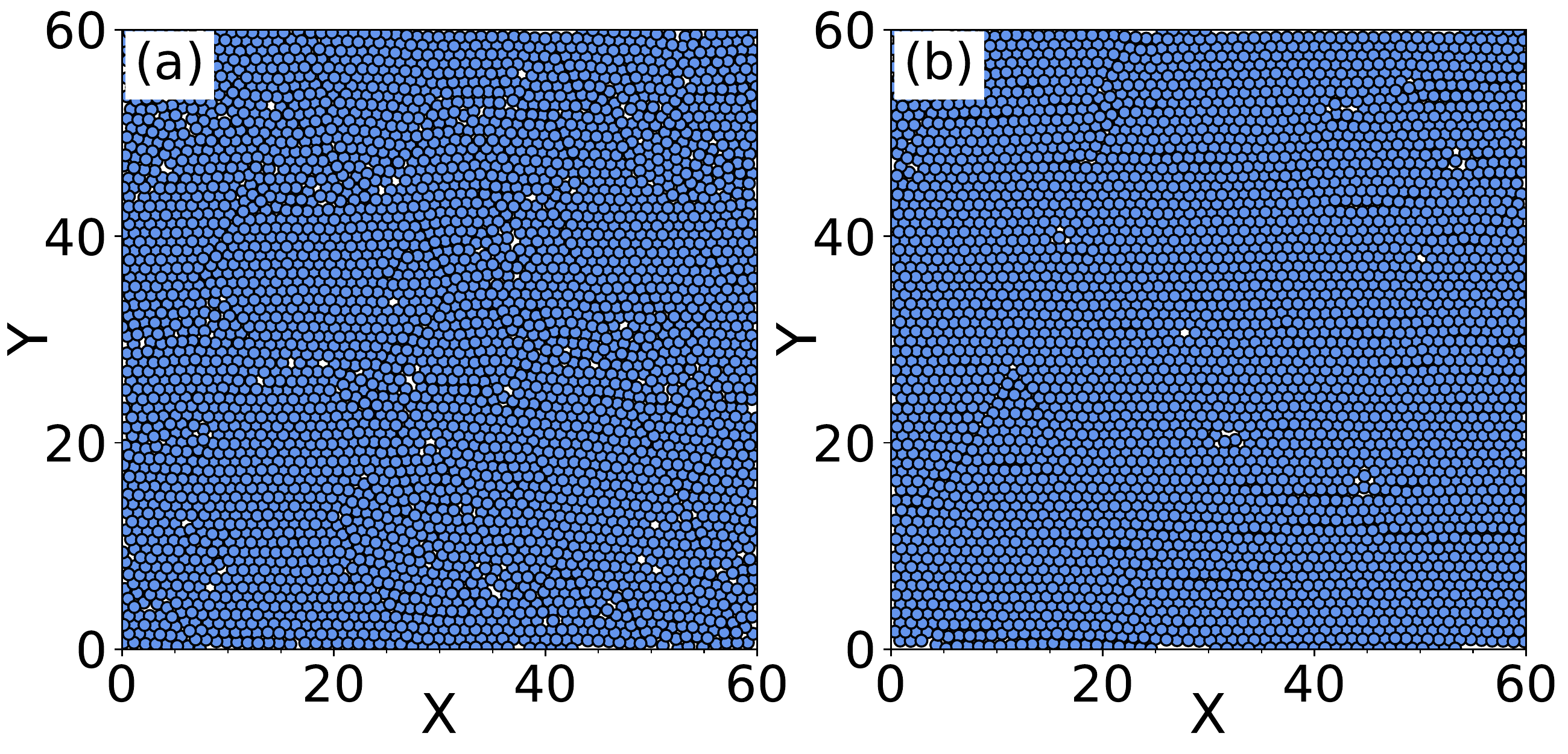}
    \caption{
      Images obtained at a disk density of $\phi =0.85$
      for the system in Fig.~\ref{fig:8},
      where the system forms a jammed solid with increasing
      triangular ordering at higher drives.
      (a) $F_D/F_p=0.25$.  (b) $F_D/F_p=2.5$.
    }
    \label{fig:9}
  \end{figure}

  For the lower and intermediate disk densities,
  there are clear changes in the particle configurations as a function of 
  drive.  At high densities of
  $\phi \geq 0.85$, however,
  the system becomes a uniform jammed solid and the
  depinning transition changes
  in character from plastic, where there can be a
  coexistence of pinned and moving particles,
  to elastic, where all the particles keep the same neighbors as they move.
  There is a distinct change in the eigenvalue distribution in Fig.~\ref{fig:4}
  for $\phi=0.85$, with significant weight appearing at higher values of $N$,
  indicating a change in the ability of the PC to capture the information in the system.
  In Fig.~\ref{fig:8} we plot $P_{1}$ versus $F_{D}/F_{P}$ for 
  $\phi = 0.85$.
  Instead of peaks, we find a
  monotonic increase in $P_1$ with increasing $F_{D}/F_{P}$. 
  In Fig.~\ref{fig:9}(a) we illustrate the disk configuration
  at $F_{D}/F_{P} = 0.25$, where the system forms
  a mostly triangular solid with a small number of vacancies.
  For $F_{D}/F_{p} =  2.5$, shown in Fig.~\ref{fig:9}(b),
  the structure is similar but the amount of triangular order is larger.
  Here the lack of jumps in $P_{1}$ is consistent with the fact that the depinning
  is elastic and the sample shows no large scale changes in the particle configurations.

  \begin{figure*}
    \includegraphics[width=0.45\textwidth]{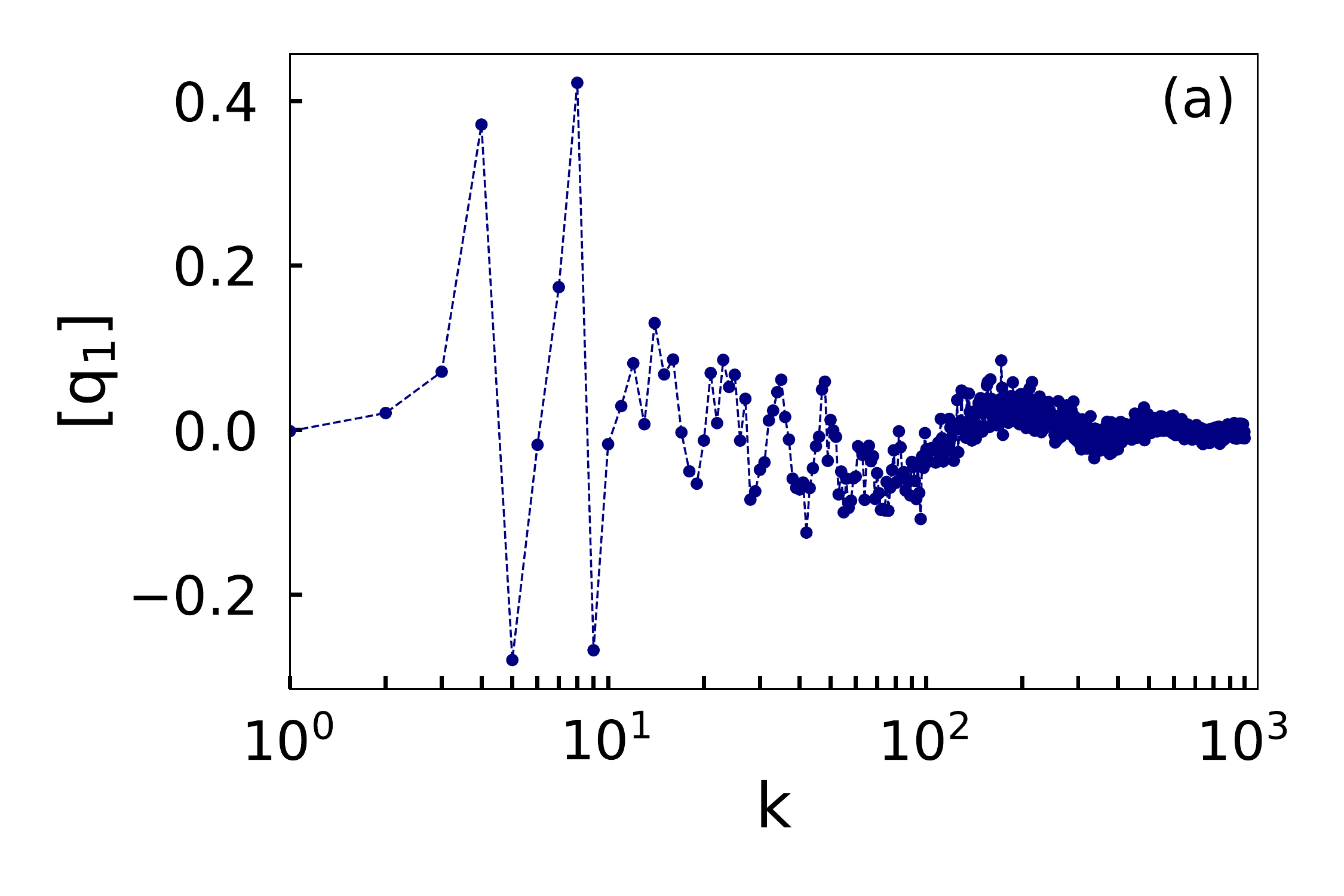} 
    \includegraphics[width=0.45\textwidth]{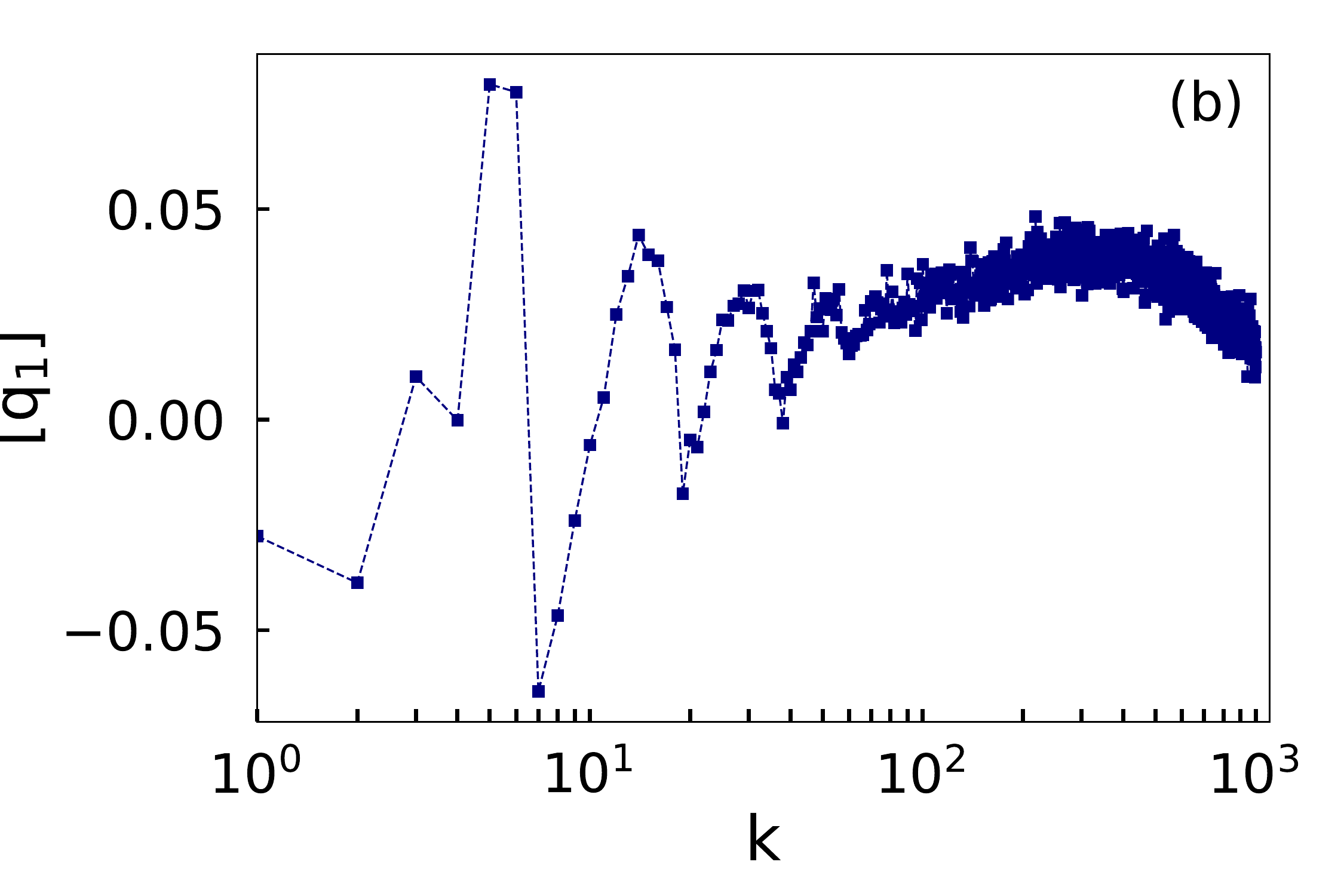} \\
    \includegraphics[width=0.45\textwidth]{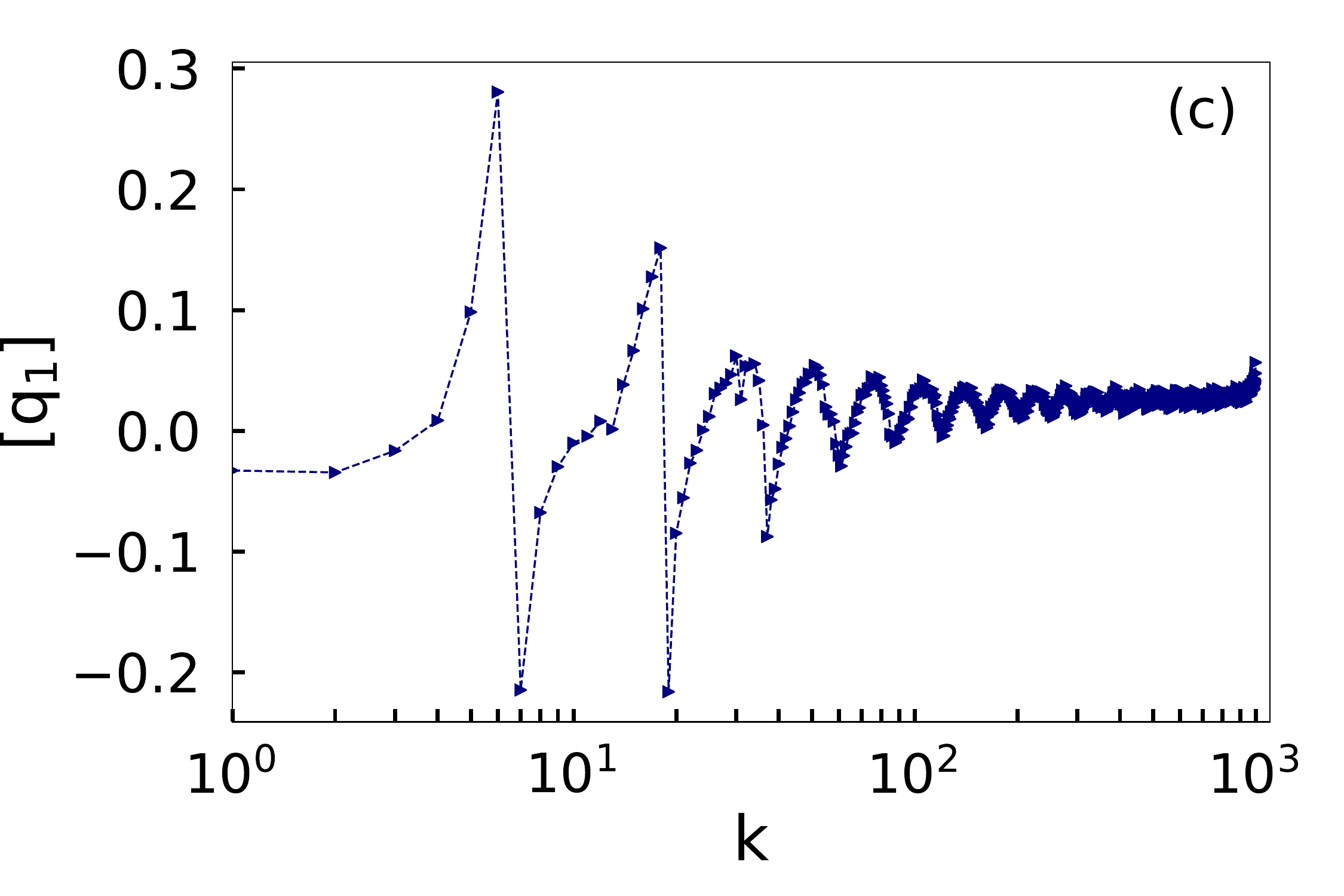} 
    \includegraphics[width=0.45\textwidth]{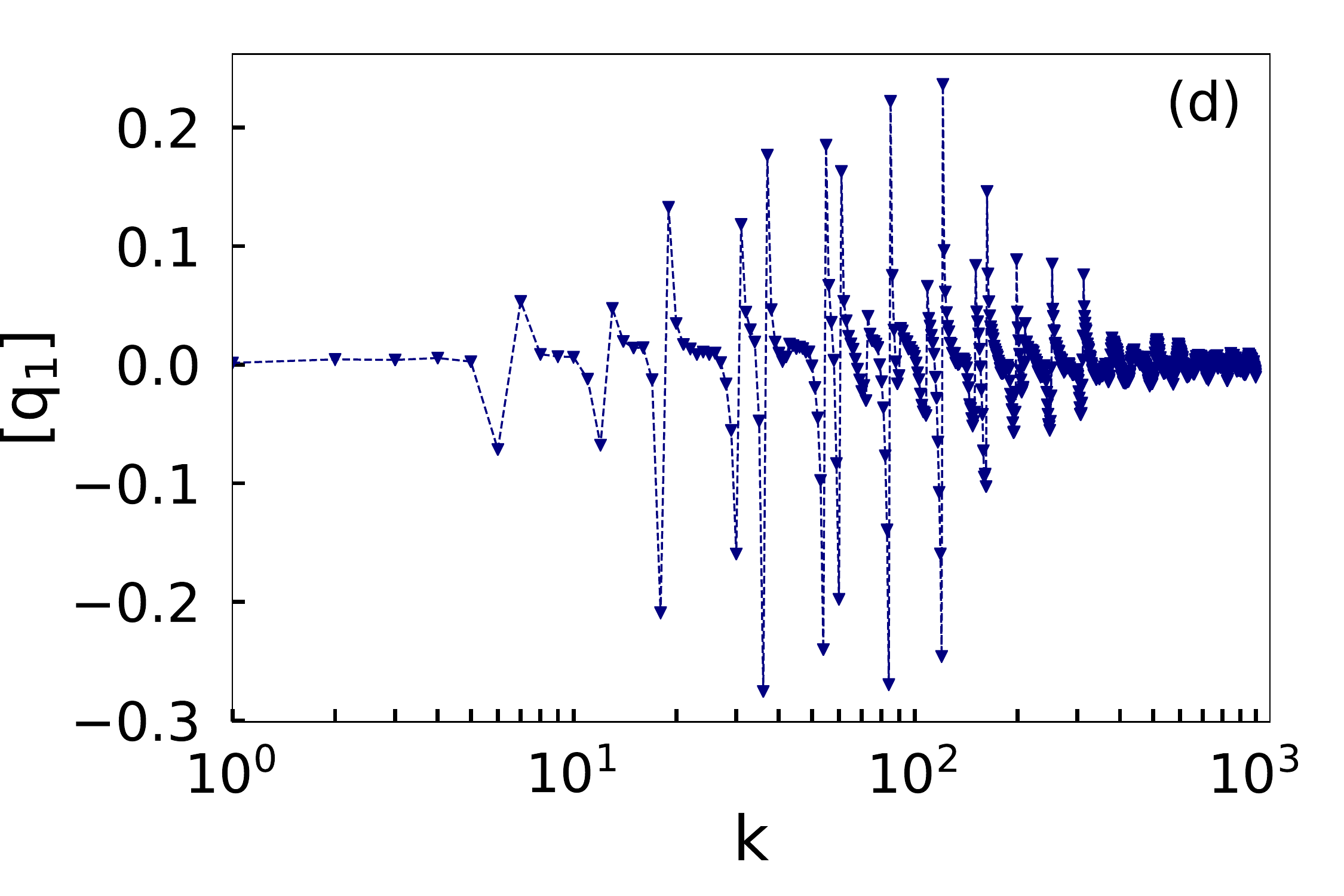}
    \caption{
      First vector $[q_1]$ from the total transformation matrix $\vec{Q}$ versus $k$,
      which is related to a neighbor distance,
      for the system in Fig.~\ref{fig:1}
      at disk densities $\phi=$ (a) 0.25, (b) 0.43, (c) 0.61, and (d) 0.85.
    }
    \label{fig:10}
  \end{figure*}
 
  The total transformation matrix $\vec{Q}=\vec{W}\vec{W}_0$
  provides a physical snapshot of the system not unlike that
  given by $g(r)$ \cite{Jadrich18,Jadrich18a}.
  When applied to a raw feature vector $\vec{f}_i$, $\vec{Q}$ first
  prewhitens the vector through the $\vec{W}_0$ matrix, and then
  transforms the vector into the PC basis through the $\vec{W}$ matrix.
  The first row of $Q$, termed
  $[q_1]$,
  is a convolution of the prewhitening transformation and the basis
  transformation for the first principal component,
  such that the expression
  $p_1=[q_1]\vec{f}_i$
  gives the mapping of the raw feature vector onto the first principal component.
  The $k$th component of $[q_1]$ provides the mapping of the $k$th component
  of $\vec{f}_i$, and since the elements of $\vec{f}_i$ are ordered according to
  neighbor distance, with the smallest values of $k$ corresponding to the smallest
  neighbor distances, it is possible to interpret $k$ as a neighbor distance.
  The prewhitening portion of $[q_1]$, plotted as a function of $k$,
  contains information similar to that
  found in $g(r)$ of an ideal gas at density $\phi$.
  The transforming portion of $[q_1]$ indicates which neighbor distances are
  most strongly weighted in the first principal component basis.
  In Fig.~\ref{fig:10} we plot
  $[q_1]$ versus $k$ 
  from a
  PCA analysis of the
  monodisperse passive disks
  at disk densities of $\phi=0.25$, 0.43, 0.61, and 0.85.
  The prewhitening component produces regular oscillations in
  $[q_1]$ at spacings corresponding to the average distance between
  successive rings of particles surrounding the probe particle.
  In an ideal gas, these oscillations would diminish with increasing $k$.
  The uneven weighting of the oscillations is an indication of which
  distance scales are important at each density in the first
  principal component.
  In Fig.~\ref{fig:10}(a-c),
  samples with low and intermediate densities of $\phi=0.25,$ $\phi=0.43,$
  and $\phi=0.61$
  have large peaks of $[q_1]$ at smaller $k$, indicating that the
  structural ordering is relatively short ranged.  In contrast,
  the high density $\phi=0.85$ sample in
  Fig.~\ref{fig:10}(d) has strong weightings at much larger $k$,
  indicating the long range nature of the emerging crystalline
  ordering in the jammed state.

  \section{Discussion}

  We have demonstrated
  that
  unsupervised machine learning can detect 
  depinning and
  the transitions between different dynamical phases in
  driven systems with quenched
  disorder.
  A similar approach could be adapted for systems
  with longer range particle-particle interactions,
  such as superconducting vortices \cite{Koshelev94,Olson98a,Kolton99},
  charged colloids, and Wigner crystals \cite{Cha94a,Reichhardt01},
  which can exhibit a pinned phase,
  plastic depinning,
  disordered liquid flow, and a moving crystal or smectic flow phase.
  In these
  systems it is often possible to use
  $P_{6}$ to detect the transitions;
  however, in some situations,
  additional transitions could be present that produce no signal
  in $P_6$ but that could be detected using PCA.
  For example,
  at a transition from a liquid to a strongly nematic
  or smectic state,
  the density of defects in the lattice undergoes little change
  and therefore the value of $P_6$ is
  constant across the transition, but the PCA
  could detect the structural change occurring in the system.
  Additionally, the plastic flow state may be composed of distinct plastic flow
  phases that have not yet been characterized but that may be detectable
  using the PCA approach.
  Depinning of particles on periodic substrates would also be
  interesting to study since in this system,
  different types of soliton or incommensurate flow patterns arise.
  These flow states can
  often be
  observed through features in the velocity-force curves,
  but produce little change in the 
  structure of the particles \cite{Reichhardt97,Bohlein12,Vanossi12,McDermott13}.
  Here, PCA could be applied to more readily distinguish 
  between the different types of commensurate and incommensurate flows. 
     
  PCA could also be applied to the class of systems
  that exhibit elastic depinning, 
  in which the particles maintain the same neighbors
  as they begin to flow \cite{Reichhardt17}. 
  Our disk system at a density of $\phi = 0.85$
  generally behaves elastically,
  keeping the same structures at depinning as in the moving
  phase, and
  we find that the PCA analysis gives distinctive results for this elastic state
  compared to the plastic flow phases that appear at lower densities.
  Elastic depinning can occur for
  superconducting vortices or
  skyrmions interacting with weak pinning,
  or in the depinning of domain walls and elastic lines. 

  Another future direction is to apply PCA to other measures beyond the
  particle configurations,
  such as velocity fluctuations, 
  the velocity-force curves, defect distributions, or local stress.
  Our results suggest that unsupervised machine learning can be a
  valuable method for identifying
  different nonequilibrium phases and the transitions between
  them. 
  Since the data we employed in our analysis
  included
  only the particle locations and not the pinning
  site locations, a similar approach could be used
  for any type of particle-based system.  
 
  \section{Summary}
  \label{sec:5}

  In summary, we have 
  shown that PCA can be used to identify
  the depinning transition and different nonequilibrium flow phases
  in a driven system of disks
  with short-range interactions moving over quenched disorder
  in the form of randomly placed pinning sites.
  In this system, traditional methods used to characterize depinning, such as the
  velocity-force curve and the Voronoi tessellation, show
  only weak signatures of the different dynamic states.
  In contrast, the PCA
  produces pronounced signals at
  the transitions
  between the pinned state, the moving phase separated
  state, and the moving smectic state.
  Using PCA, we also find evidence for
  more subtle phase transitions such as a clustered pinned phase as well as
  a transition between an amorphous and a crystalline phase separated state.
  The PCA can detect the onset of the jammed state and
  exhibits
  different signatures
  for plastic versus elastic depinning. The PCA
  method can be used to
  search for additional features in previously
  studied depinning systems such as superconducting vortices, Wigner crystals,
  skyrmions, and charge density wave systems,
  as well as to identify novel nonequilibrium phases in particle-based systems.

  \acknowledgments
  This work was supported by the US Department of Energy through
  the Los Alamos National Laboratory.  Los Alamos National Laboratory is
  operated by Triad National Security, LLC, for the National Nuclear Security
  Administration of the U. S. Department of Energy (Contract No.~892333218NCA000001).
  This research was supported in part by
  the M. J. Murdock Charitable Trust and the
  Notre Dame Center for Research Computing.

\bibliography{mybib}
\end{document}